\documentclass[12pt]{article}
\usepackage[english]{babel}
\usepackage{graphicx,epsfig}
\makeindex \textwidth 170mm \textheight 220mm \topmargin -5mm
\newcommand{\be}{\begin{equation}}
\newcommand{\ee}{\end{equation}}
\newcommand{\bea}{\begin{eqnarray}}
\newcommand{\eea}{\end{eqnarray}}

\def\bb{\bibitem}
\def\eqi{\begin{equation}}
\def\eqf{\end{equation}}
\def\eqia{\begin{eqnarray}}
\def\eqfa{\end{eqnarray}}
\def\btab{\begin{tabular}}
\def\etab{\end{tabular}}
\def\bar{\begin{array}}
\def\ear{\end{array}}

\def\leti{Lense--Thirring}

\def\lg{{\rm LAGEOS}}
\def\lgg{{\rm LAGEOS} II}

\def\bm#1{{\mbox{\boldmath$#1$\unboldmath}}}

\def\zone{the error due to the even zonal harmonics of the geopotential\ }

\def\rp#1#2{{#1\over#2}}
\def\msy{mas yr$^{-1}$}
\begin{document}
\begin{titlepage}
\begin{flushright}
\today\\
BARI-TH/00\\
\end{flushright}
\vspace{.5cm}
\begin{center}
{\LARGE A reassessment of the systematic gravitational error in
the LARES mission} \vspace{1.0cm}
\quad\\
{Lorenzo Iorio$^{\dag}$\\ \vspace{0.1cm}
\quad\\
{\dag}Dipartimento di Fisica dell' Universit{\`{a}} di Bari, via
Amendola 173, 70126, Bari, Italy}\\ \vspace{0.2cm} \vspace{0.2cm}
\vspace{1.0cm}

{\bf Abstract\\}
\end{center}

{\noindent \small  In this letter we reexamine the evaluation of
\zone in some proposed tests of relativistic gravitomagnetism with
existing and proposed laser--ranged LAGEOS--like satellites in the
gravitational field of the Earth. It is particularly important
because \zone is one of the major sources of systematic errors in
this kind of measurements. A conservative, although maybe
pessimistic, approach is followed by using the diagonal part only
of the covariance matrix of the EGM96 Earth's gravity model up to
degree $l=20$. It turns out that, within this context and
according to the present level of knowledge of the terrestrial
gravitational field, the best choice would be the use of a
recently proposed combination which involves the nodes $\Omega$ of
LAGEOS, LAGEOS II and LARES and the perigees $\omega$ of LAGEOS II
and LARES. Indeed, it turns out that the unavoidable orbital
injection errors in the inclination of LARES would induce a
gravitational error which turns out to be insensitive both to the
even zonal harmonics of degree higher than $l=20$ and to the
correlation among them.}
\end{titlepage} \newpage \pagestyle{myheadings} \setcounter{page}{1}
\vspace{0.2cm} \baselineskip 14pt

\setcounter{footnote}{0}
\setlength{\baselineskip}{1.5\baselineskip}
\renewcommand{\theequation}{\mbox{$\arabic{equation}$}}
\noindent

\section{Introduction}
One of the most intriguing  prediction of the General Theory of
Relativity, in its linearized weak--field and slow--motion
approximation, is the so called frame--dragging or Lense--Thirring
effect (Lense and Thirring 1918). It can be thought of as a
consequence of a gravitational coupling between the proper angular
momentum \bm J of a central body of mass $M$, which acts as source
of the gravitational field, and the angular momentum \bm s of a
particle freely orbiting it. It turns out that the spin \bm s
undergoes a tiny precessional motion (Schiff 1960). The most
famous experiment devoted to the measurement of such a
gravitomagnetic effect is the Stanford University GP--B mission
(Everitt $et\ al $ 2001) which is scheduled to fly in April 2003.

If we consider the whole orbit of a test particle in its geodesic
motion around $M$ as a sort of giant gyroscope, its orbital
angular momentum \bm l undergoes the Lense--Thirring precession,
so that the longitude of the ascending node $\Omega$ and the
argument of pericenter $\omega$ of the orbit of the test particle
(Sterne 1960) are affected by small secular
precessions\footnote{In the original paper by Lense and Thirring
the longitude of the pericenter $\varpi=\Omega+\omega$ is used
instead of $\omega$. } $\dot\Omega_{\rm LT}$, $\dot\omega_{\rm
LT}$ (Lense and Thirring 1918; Ciufolini and Wheeler 1995; Iorio
2001).
\subsection{The LAGEOS-LAGEOS II Lense-Thirring experiment}
Up to now, the only attempts to detect them in the gravitational
field of the Earth are due to Ciufolini and coworkers (Ciufolini
$et\ al$ 1998; Ciufolini 2000; 2002) who analysed the laser data
of the existing geodetic SLR (Satellite Laser Ranging) satellites
LAGEOS and LAGEOS II over time spans of some years. The observable
is a suitable combination of the orbital residuals of the nodes of
LAGEOS and LAGEOS II and the perigee of LAGEOS II according to an
idea exposed in (Ciufolini 1996). The relativistic signal is a
linear trend with a slope of almost 60.2 milliarcseconds per year
(\msy\ in the following). The claimed total accuracy is of the
order of $20\%-30\%$. The main sources of systematical errors in
such kind of measurements are the unavoidable aliasing effect due
to the mismodelling in the classical secular precessions induced
by the even zonal coefficients of the multipolar expansion of the
Earth's gravitational field (Kaula 1966) and the
non--gravitational perturbations affecting especially the perigee
of LAGEOS II. It turns out that the mismodelled classical
precessions due to the first two even zonal harmonics of the
geopotential $J_2$ and $J_4$ are the most insidious source of
error for the Lense--Thirring measurement with LAGEOS and LAGEOS
II. The combination of (Ciufolini 1996) is insensitive just to
$J_2$ and $J_4$. According to the full covariance matrix of the
EGM96 gravity model (Lemoine $et\ al$ 1998), the error due to the
remaining uncancelled even zonal harmonics amounts to almost
13$\%$. A reliable evaluation of the impact of the geopotential is
a particularly subtle and important topic. Indeed, it is based on
the use of the covariance matrix of the even zonal harmonics of
the geopotential of some terrestrial gravity models like EGM96. As
pointed out in (Ries $et\ al$ 1998), in obtaining the solution of
EGM96 and of other previous gravity models a multidecadal
observational time span has been used and many seasonal,
stochastic and secular variations, which is known that they affect
the geopotential, have not been accounted for. Then, according to
the remarks of (Ries $et\ al$ 1998), nothing would assure that
during any particular relatively short time span as that used in
the LAGEOS--LAGEOS II Lense--Thirring experiment the correlation
among the geopotential coefficients would be just that of the
EGM96 covariance matrix. For example, it seems that the $13\%$
favorable estimate is based on a particular correlation between
$J_6$ and $J_8$ (Ries $et \ al$ 1998) which do affect the
observable by Ciufolini. If, with a more conservative, although
maybe pessimistic, approach, we use the diagonal part only of the
covariance matrix of EGM96 up to degree $l=20$ \zone for the
LAGEOS--LAGEOS II Lense--Thirring experiment amounts to  $46.6\%$
(Iorio 2002a). Moreover, it turns out to be insensitive to the
even zonal harmonics of degree higher than $l=20$ due to the high
altitude of the LAGEOS satellites. Indeed, the classical
precessions of the node and the perigee depend on
$\left(\frac{R}{a}\right)^l a^{-\rp{3}{2}}$, where $R$ is the
Earth's radius and $a$ is the satellite's semimajor axis.
\subsection{The LARES project}
The originally proposed LARES mission (Ciufolini 1986; 1998)
consists of the launch of a LAGEOS--type satellite--the
LARES--with the same orbit of LAGEOS except for the inclination
$i$ of its orbit, which should be supplementary to that of LAGEOS,
and the eccentricity $e$, which should be one order of magnitude
larger. In Table 1 the orbital parameters of the existing and
proposed LAGEOS--type satellites are quoted.
\begin{table}[ht!]
\caption{Orbital parameters of \lg, \lgg, LARES and POLARES and
their Lense--Thirring precessions.} \label{tavola1}
\begin{center}
\begin{tabular}{llllll}
\noalign{\hrule height 1.5pt} Orbital parameter & \lg & \lgg &
LARES & POLARES\\ \hline
$a$ semi major axis (km) & 12,270 & 12,163 & 12,270 & 8,378\\
$e$ eccentricity & 0.0045 & 0.014 & 0.04 & 0.04\\
$i$ inclination (deg) & 110 & 52.65 & 70 & 90\\
$\dot\Omega_{\rm LT}$ (\msy) & 31 & 31.5 & 31 & 96.9\\
$\dot\omega_{\rm LT}$ (\msy)& 31.6 & -57 & -31.6 & 0\\
\noalign{\hrule height 1.5pt}
\end{tabular}
\end{center}
\end{table}
The choice of the particular value of the inclination for the
LARES is motivated by the fact that in this way, by using as
observable the sum of the nodes of the LAGEOS and the LARES, it
should be possible to cancel out exactly all the contributions of
the even zonal harmonics of the geopotential, which depends on
$\cos i$, and add up the Lense--Thirring precessions which,
instead, are independent of $i$.

Of course, it would not be possible to obtain practically two
orbital planes exactly 180 deg apart due to the unavoidable
orbital injection errors. It turns out that all depends on the
last stadium of the rocket used. According to conservative
estimates, if a solid propellant is used for it an error in
inclination of the order of 1 deg is to be expected, while if a
liquid propellant, which is more expensive, is used  the error
should amount to 0.5-0.6 deg (Anselmo, private communication
2002). In Figure 1, page 4314 of (Iorio $et \ al$ 2002) the impact
of such source of error on the originally proposed LAGEOS--LARES
mission has been shown. It should be noted that the simple sum of
the nodes of the LAGEOS and the LARES is affected by all the even
zonal harmonics of the geopotential. Then, when the impact of the
departures of the inclination of LARES from its nominal values on
\zone has to be calculated, the role of all the correlations among
the even zonal harmonics should be important. If we decide to take
into account the remarks of (Ries $et\ al$ 1998), the results
obtained in (Iorio $et\ al$ 2002) might be considered optimistic
in the sense that they are based on an extrapolation of the
validity of the full covariance matrix of EGM96 up to degree
$l=20$ to arbitrary future time spans.

In (Iorio $et\ al$ 2002) an alternative observable based on the
combination of the residuals of the nodes of LAGEOS, LAGEOS II and
LARES and the perigee of LAGEOS II and LARES has been proposed. It
would allow to cancel out the first four even zonal harmonics
$J_2,\ J_4,\ J_6,\ J_8$ so that \zone would be rather insensitive
to the orbital injection errors in the LARES inclination and would
amount to $0.02\%$ only.

In (Iorio 2002b) the recent proposal of inserting the LARES in a
low--altitude polar orbit (Lucchesi and Paolozzi 2001), so to
obtain the so called POLARES, and to analyze only its node has
been critically analyzed from the point of view of the impact of
the orbital injection errors in the POLARES inclination.
\subsection{Motivation of the present work}
The conclusions obtained in (Iorio $et\ al$ 2002; Iorio 2002b) are
based on the assumption of the validity of the EGM96 full
covariance matrix in arbitrary future time spans during which,
instead, it might happen that the correlations between the even
zonal harmonics will be different. This problem is particularly
relevant for those observables which are sensitive to the full
range of the even zonal harmonics of the geopotential like the
originally proposed node--only LAGEOS--LARES combination and the
node--only POLARES observable.

Consequently, we wish to reanalyze such issues in a more
conservative, although pessimistic, approach by using the diagonal
part only of the covariance matrix of EGM96. We expect that \zone
of the new proposed observable based on the use of the orbital
elements of LAGEOS, LAGEOS II and LARES, which is $J_2-J_8$-free,
should be relatively insensitive to the correlation among the
remaining even zonal harmonics, contrary to the sum of the nodes
of the LAGEOS and the LARES and the node of the POLARES. If it
will be so, the reliability of the modified version of the LARES
project will be enforced and posed on a more firm basis.
\section{The LARES mission}
\subsection{The originally proposed LARES scenario}
As pointed out in (Iorio $et \ al$ 2002), the impact of the
unavoidable orbital injection errors in the LARES inclination on
\zone is of crucial importance for the originally proposed
LAGEOS--LARES observable, especially if the LARES satellite will
be finally launched with a relatively cheap rocket of not too high
quality due to budget restrictions. Figure 1 of page 4314 in
(Iorio $et \ al$ 2002) has been obtained by considering the
root--sum--square error due to the full covariance matrix of
EGM96, up to degree $l=20$, as a function of the inclination of
LARES. Since there are no even zonal harmonics cancelled out by
the sum of the nodes of LAGEOS and LARES, the role of the
correlation among all the various geoptential's harmonics in the
assessment of \zone should not be neglected. Then, the estimates
of (Iorio $et \ al$ 2002) might reveal to be rather optimistic. A
more conservative approach consists of repeating the analysis by
using the diagonal part only of the covariance matrix of EGM96.
The results are summarized in Figure 1.
%%%%%%%%%%%%%%%%%%%%%%%%%%%%%%%%%%%%%%%%%%%%%%%%%%%%%%%%%%%%%%%%%%%%%%%%%%%%%%%%%%
\begin{figure}[ht!]
\begin{center}
\includegraphics*[width=13cm,height=10cm]{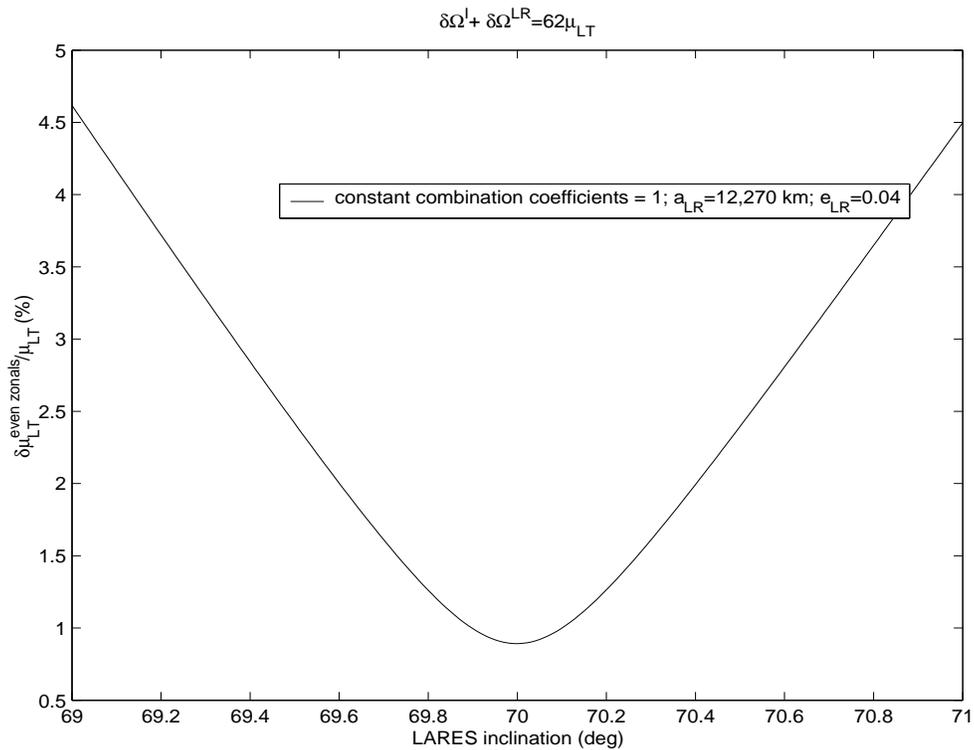}
\end{center}
\caption{\footnotesize Influence of the injection errors in the
LARES inclination on the error of the originally proposed
LAGEOS--LARES nodal observable due to the even zonal harmonics of
the geopotential according to the diagonal part only of the
covariance matrix of EGM96 up to degree $l=20$.} \label{figura1}
\end{figure}
%%%%%%%%%%%%%%%%%%%%%%%%%%%%%%%%%%%%%%%%%%%%%%%%%%%%%%%%%%%%%%%%%%%%%%%%%%%%%%%%
It can be noticed that, in this pessimistic but perhaps more
realistic approach, \zone is of the order of 4$\%$--4.5$\%$,
contrary to the 1$\%$--1.4$\%$ of (Iorio $et \ al$ 2002). Note
also that, even for the nominal values of Table 1 for the LARES
orbit, \zone would amount to almost 1$\%$, contrary to 0.3$\%$ of
(Iorio $et \ al$ 2002). This further confirms that, according to
the present knowledge of the terrestrial gravitational field, the
implementation of the originally proposed LAGEOS--LARES observable
would pose some problems in term of accuracy. Of course, the
situation should greatly improve when the new data for the
geopotential from the CHAMP and GRACE missions will be available.
It should also be considered that, in the case of the node--only
LAGEOS--LARES configuration, \zone would represent the most
relevant part of the systematic error because the
non--gravitational perturbations acting on the nodes of the
LAGEOS--like satellites are far less relevant.
\subsection{The modified LARES scenario}
The combination of orbital residuals including the nodes of
LAGEOS, LAGEOS II and LARES and the perigees of LAGEOS II and
LARES of eq. (9) in (Iorio $et \ al$ 2002) seems to be a better
choice. Indeed, also in this pessimistic approach \zone turns out
to be very small and insensitive to the orbital injection errors
in the inclination of the LARES satellite, as shown in Figure 2.
%%%%%%%%%%%%%%%%%%%%%%%%%%%%%%%%%%%%%%%%%%%%%%%%%%%%%%%%%%%%%%%%%%%%%%%%%%%%%%%%%%%%
\begin{figure}[ht!]
\begin{center}
\includegraphics*[width=13cm,height=10cm]{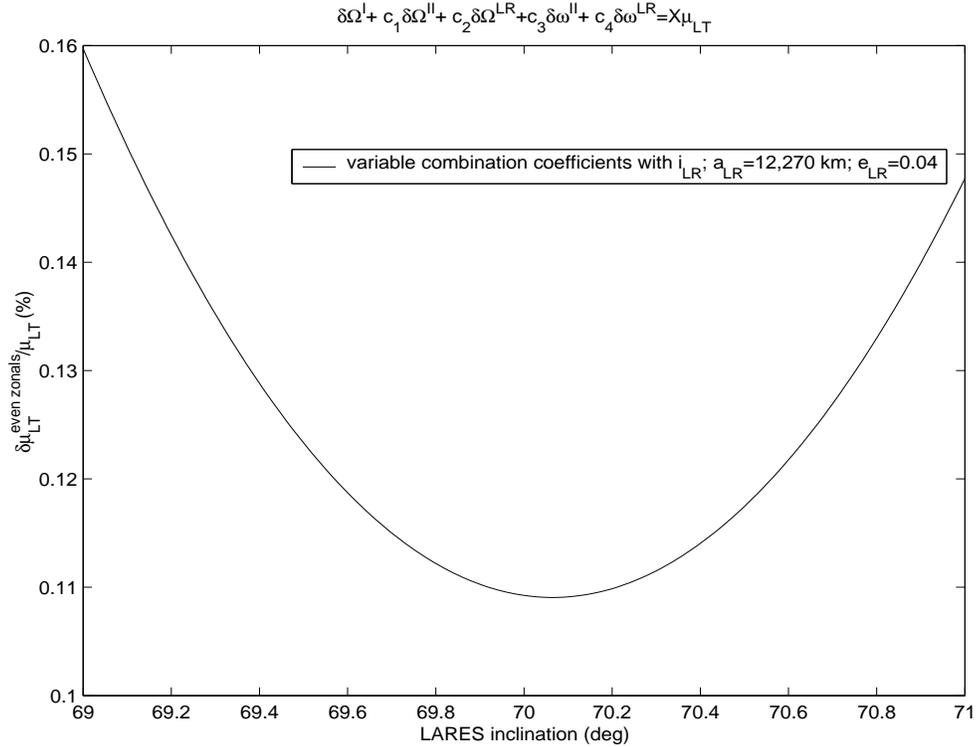}
\end{center}
\caption{\footnotesize Influence of the injection errors in the
LARES inclination on the error of the LAGEOS--LAGEOS II--LARES new
observable due to the even zonal harmonics of the geopotential
according to the diagonal part only of the covariance matrix of
EGM96 up to degree $l=20$.} \label{figura2}
\end{figure}
%%%%%%%%%%%%%%%%%%%%%%%%%%%%%%%%%%%%%%%%%%%%%%%%%%%%%%%%%%%%%%%%%%%%%%%%%%%%%555
From it \zone remains almost fixed at the level of 0.1$\%$, while
in Figure 3 of page 4317 in (Iorio $et \ al$ 2002) it is of the
order of 0.02$\%$--0.04$\%$. As it could be expected, since the
first four even zonal harmonics are cancelled out by the combined
residuals, the impact of the correlation among the remaining ones
is relatively not relevant.

Notice also that the part of the systematic error due to the
non--gravitational perturbations, as evaluated in (Iorio $et \ al$
2002) over a time span of 7 years, amounts just to 0.3$\%$. This
is very important because this means that the combined residuals
approach would yield a more accurate measurement of the
Lense--Thirring effect than the simple sum of the nodes of the
LAGEOS and LARES satellites, even with the conservative
evaluations of \zone presented here. Indeed, for, say, $\delta
i^{\rm inj}_{\rm LARES}\sim 1$ deg, the total systematic error of
the combination of residuals, according to Figure 2, would amount
to 0.33$\%$, while for the sum of the nodes it would be of the
order of 4.5$\%$, even if the impact of the non--gravitational
perturbations on the nodes is very small and is considered
negligible.

Also in this case the role which will be played by the results of
the CHAMP and GRACE missions will be of decisive importance.
\section{The POLARES scenario}
The approach followed in this letter for \zone clearly shows in
Figure 3 that the option of inserting the LARES satellite in a
polar, low--altitude orbit should be considered rather
impracticable.
%%%%%%%%%%%%%%%%%%%%%%%%%%%%%%%%%%%%%%%%%%%%%%%%%%%%%%%%%%%%%%%%%%%%%%%%%%%%%%%%%%%%%5
\begin{figure}[ht!]
\begin{center}
\includegraphics*[width=13cm,height=10cm]{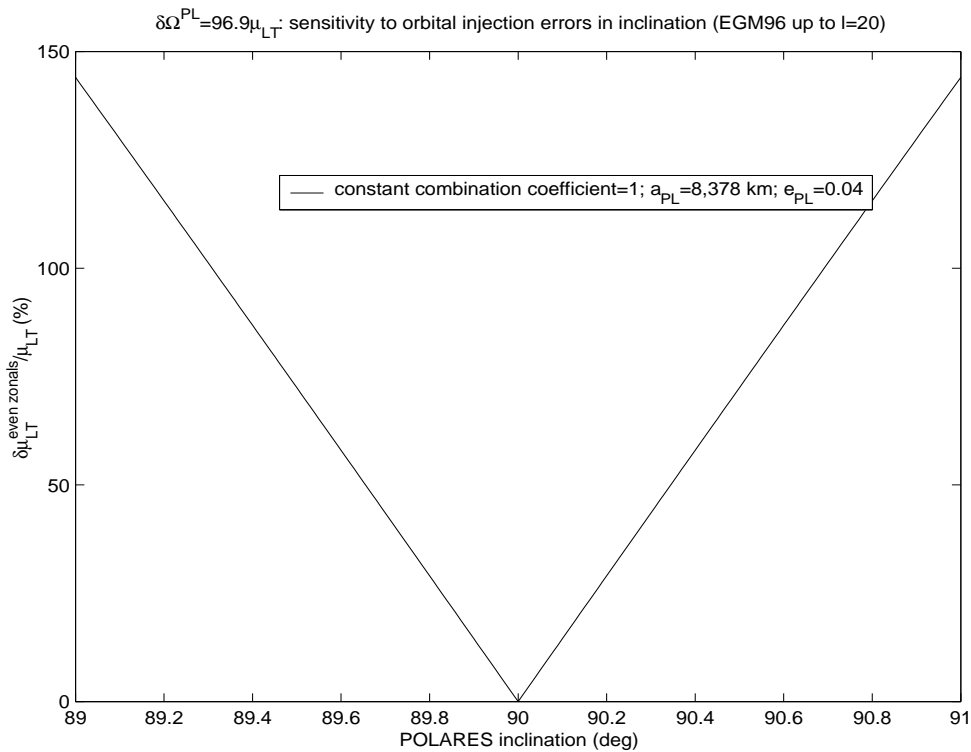}
\end{center}
\caption{\footnotesize Influence of the injection errors in the
POLARES inclination on the error of the POLARES node due to the
even zonal harmonics of the geopotential according to the diagonal
part only of the covariance matrix of EGM96 up to degree $l=20$.}
\label{figura3}
\end{figure}
%%%%%%%%%%%%%%%%%%%%%%%%%%%%%%%%%%%%%%%%%%%%%%%%%%%%%%%%%%%%%%%%%%%%%%%%%%%%%%%%%%%
Indeed, the effects of the orbital injection errors in the
inclination of POLARES would be greatly enhanced by its low
altitude of 2000 km only at a level of 100$\%$--150$\%$. It should
also be considered that the results of Figure 3, as those of
Figure 1 of page L178 in (Iorio 2002b), are optimistic because
they have been obtained by neglecting the contributions of the
classical nodal precessions of the even zonal harmonics of degree
higher than $l=20$ which, contrary to the LAGEOS satellites, would
not be negligible in this case due to the low altitude of the
POLARES. Moreover, it is not probable that the improvements due to
the CHAMP and GRACE missions will reduce \zone to a level
comparable to the other proposed configurations, which, in turn,
will benefit from the new Earth gravity models.
\section{Conclusions}
In this letter we have followed a more conservative and realistic
approach in evaluating the impact of the orbital injection errors
of LARES and POLARES on \zone of some proposed observables aimed
to the measurement of the Lense--Thirring effect with LAGEOS--like
SLR satellites. In particular, we have used, in a
root--sum--square fashion, the diagonal part only of the
covariance matrix of the even zonal harmonic coefficients of the
EGM96 Earth's gravity model up to degree $l=20$.

With regard to the LARES project, it turns out that the recently
proposed residuals combination involving the LAGEOS, LAGEOS II and
LARES satellites, which cancels out the first four even zonal
harmonics, should yield not only a more accurate measurement than
the simple sum of the nodes of LAGEOS and LARES, but also more
reliable because it would be less dependent on the correlation
between the remaining even zonal harmonics of higher degree.

The use of POLARES, according to the present level of knowledge of
the terrestrial gravitational field and to the results presented
here, should be considered unpracticable. It is probable that,
even with the new gravity models from CHAMP and GRACE, such a
proposed configuration would not be competitive with the other
multisatellite observables which, in turn, would benefit of the
new data of the gravitational field.

The new gravity models from the CHAMP and GRACE missions should
yield great benefits for a more confident and reliable assessment
of \zone in all the examined missions. However, also in this case
the originally proposed LARES orbital configuration, in conjuction
with the data from LAGEOS and LAGEOS II, should be the best
choice.
%-----------------------------------------
\section*{Acknowledgements}
L.I. is grateful to L. Guerriero for his support while at Bari.
%-----------------------------------------

\end{document}